\documentclass[12pt,preprint]{aastex}

\begin{document}



\title{On how much X-ray and UV radiation processes are coupled 
in accretion disks: AGN case}

\author{Daniel Proga,\altaffilmark{1}}

\affil{$^1$ Princeton University Observatory, Peyton Hall, Princeton, NJ 08544.e-mail: dproga@astro.princeton.edu}

\def\LSUN{\rm L_{\odot}}
\def\MSUN{\rm M_{\odot}}
\def\RSUN{\rm R_{\odot}} 
\def\MSUNYR{\rm M_{\odot}\,yr^{-1}}
\def\MSUNS{\rm M_{\odot}\,s^{-1}}
\def\MDOT{\dot{M}}

\newbox\grsign \setbox\grsign=\hbox{$>$} \newdimen\grdimen \grdimen=\ht\grsign
\newbox\simlessbox \newbox\simgreatbox
\setbox\simgreatbox=\hbox{\raise.5ex\hbox{$>$}\llap
     {\lower.5ex\hbox{$\sim$}}}\ht1=\grdimen\dp1=0pt
\setbox\simlessbox=\hbox{\raise.5ex\hbox{$<$}\llap
     {\lower.5ex\hbox{$\sim$}}}\ht2=\grdimen\dp2=0pt
\def\simgreat{\mathrel{\copy\simgreatbox}}
\def\simless{\mathrel{\copy\simlessbox}}

\begin{abstract}
Within the standard accretion disk theory for active galactic nuclei (AGN), 
the observed X-rays are often modeled by Compton up-scattering 
of ultraviolet (UV) disk photons inside a hot disk corona.
Here, we point out that for many AGN, radiation pressure due 
to the very same UV disk photons can drive a flow from the disk into 
the corona and couple the processes producing X-rays and UV photons.
This coupling could lead to quenching of the disk corona 
because the regions above the UV disk will be too dense, too opaque, 
and consequently too cold. We discuss various consequences of 
this new type of the X-ray/UV coupling on the dynamical and 
radiative properties of AGN. 
\end{abstract}


\section{Introduction}
High X-ray fluxes observed in active galactic nuclei (AGN) 
are a serious challenge for the standard accretion disk theory.
They are also a serious challenge for any AGN outflow model
because no matter what is the physics of
X-ray production, one has to deal with the so-called overionization 
problem:  how outflows absorbing the ultraviolet (UV) radiation avoid 
full photoionization due to strong  X-ray radiation.

In most pictures depicting the centers of AGN, the X-rays are 
produced in a small region, referred to as the central engine, whereas 
the outflows are produced outside of the central engine. 
Therefore, the X-ray production mechanisms
are usually considered to operate separately from the outflow production 
mechanisms.
In this paper, we discuss new dynamical and radiative consequences of 
UV  photons being emitted into the central engine.
In particular, we argue that radiation pressure due to UV lines 
(line force) couples the X-ray and UV radiation processes 
by driving the disk material into the regime above the disk
where X-rays are to be produced.

\section{Disk Corona and Disk Outflow: ``a tale of two merging cities''} 
The spectral energy distribution (SED) of active galactic nuclei (AGN) 
is very broad. It spans the wavelength range from radio to hard X-rays.
Most of the AGN luminosity, $L$ is in 
the optical--UV--IR regime but some significant fraction is in the X-ray band. 
It is commonly accepted that AGN are powered by accretion of matter 
onto a supermassive black hole (BH). AGN are typically very luminous 
(0.001~$L_{Edd}\simless L \simless 1~L_{Edd}$, where $L_{Edd}$ 
is the Eddington limit).
For most AGN, the accretion flow
is thought to form an optically thick, geometrically thin Keplerian
disk. In the standard picture, this disk radiates thermally mostly
in the optical--UV regime (Shakura \& Sunyaev 1973).

\subsection{Disk Corona}
Generally, the optically thick disk model can account for
the optical--UV radiation but it does not account for the spectral shape
and high flux observed in X-rays. 
The most commonly accepted model for the production of X-rays is 
multiple Compton up-scattering (Comptonization) of UV photons 
(e.g., Sunyaev \& Titarchuck 1980) off hot electrons in a disk corona
(e.g., Walter \& Courvoisier 1992; Haardt \& Maraschi 1991, 1993;
Sobolewska et al. 2004).
The derived X-ray spectrum depends on 
the temperature and optical
depth of the scattering electrons.
Despite recent advances in observations and modeling 
of X-rays, the geometry and radiation processes responsible for X-ray
emission remain poorly constrained. 
Two types of geometries are being considered for an accretion disk
and Comptonizing corona: (1) ``slab'' or ``sandwich'' geometries 
(e.g., the top and bottom panels of  Fig.~1 in Reynolds \& Nowak 2003, 
hereafter RN) and
(2) ``sphere+disk geometries'' (e.g., the middle two panels  
of  Fig.~1 in RN). 

Here we focus on exploring the first class of geometries where
the hot corona is thought to be located immediately above the disk.
The disk emits soft thermal photons which provide 
the main source of cooling for the hot electrons in the corona.
It is possible that at the same time, the hard photons produced by
Comptonization are an important source of heating for the reflecting
matter which reprocesses them into soft photons.
Thus there is a radiative coupling between the X-ray and UV radiation.
Many models account for this coupling and assume 
that a significant fraction of the gravitational power
is dissipated in the hot tenuous layers  
(e.g., Haardt \& Maraschi 1993; Svensson \& Zdziarski 1994). 
Although the basic idea behind
a hot disk corona is  straightforward, the physical model of the corona
remains one of the biggest challenges in the field. Phenomenologically,
one can imagine that the corona is heated by dissipation of
the accretion power via magnetic processes (e.g., Galeev et al. 1979; 
Field \& Rogers 1993).

Direct studies of these complex multidimensional, time-dependent  
processes by means of numerical magnetohydrodynamical (MHD) simulations 
have begun. For example, Miller \& Stone (2000) found that in 
a stratified disk, a  MHD turbulence driven by the magnetorotational 
instability (MRI, Balbus \& Hawley 1998) is capable of driving 
magnetogravitational modes of the Parker instability.
However, it has not been demonstrated yet that magnetic buoyancy 
can supply the disk corona with sufficient power to explain 
the observed X-ray emission. In fact, several numerical simulations indicate 
that local dissipation not magnetic buoyancy
is the primary saturation mechanism of the MRI (e.g., Brandenburg et al. 1995; 
Stone et al. 1996; Miller \& Stone 2000, Turner 2004).

\subsection{Disk Wind}
Viable models for AGN must also account for
mass outflows which is another important aspect of activity in galactic nuclei.
The most relevant to this paper are the outflows that can be 
inferred from spectral features observed in the UV and X-ray bands
(i.e., we do not discuss AGN jets).
Broad absorption lines (BALs) in QSOs are the best example
of spectral features revealing the existence of
such outflows. These lines are almost always blueshifted relative to 
the emission-line rest frame, indicating the presence of outflows from 
the active nucleus, with velocities as large as 0.2~c
(e.g., Turnshek 1998). The apparent X-ray properties can be affected
by outflows, too. For example, 
the relative strength of the soft X-ray flux anti-correlates 
with the C~IV absorption equivalent width for QSOs 
(e.g., Brandt, Laor \& Wills 2000). 

There has been considerable time and effort spent to understand AGN outflows
(e.g., Arav, Shlosman \& Weymann 1997; Crenshaw, Kraemer \& George 2002, 
and references therein). In particular, many theoretical models have been 
proposed to explain outflows in AGNs 
(e.g., Crenshaw et al. 2002).  
One of most plausible scenarios for AGN outflows is a wind from an
accretion disk around a black hole where
the line force drives a wind from a disk by the 
local disk radiation at radii where the disk radiation is mostly in the UV
(e.g., Murray et al. 1995,  MCGV hereafter; 
Proga, Stone, \& Kallman 2000, PSK hereafter; 
Proga \& Kallman 2004, PK hereafter).
In this scenario, AGN outflows are a natural consequence of the standard 
accretion disk theory because the theory predicts high enough radiative flux 
and gas opacity in the UV regime, for the line force to drive an outflow. 

Numerical simulations of the radiation driven disk winds
illustrate why and how a wind from an
accretion disk can account for AGN outflows. In particular, the
simulations have been essential in 
studying the robustness of the radiation launching and
acceleration of the wind (e.g., PSK, PK). 
For example, PK considered 
relatively unfavorable conditions for line driving (LD) as they took into 
account the central engine radiation as a source of ionizing photons 
but neglected its contribution to the radiation force. Additionally,
they accounted for the attenuation of the X-ray radiation by computing
the X-ray optical depth in the radial direction assuming that only electron 
scattering contributes to the opacity. The main result of the simulations 
is that the disk atmosphere can 'shield' itself  from external X-rays
so that the local disk radiation can launch gas off the disk
photosphere.  

\subsection{The Role of the Failed Wind}

LD can change the flow near the accretion disk 
in many ways. A powerful wind is just the most dramatic of them 
for which fairly strict requirements must be met. 
Using some physical arguments as well as 
numerical simulations one can show that line-driven disk winds 
are produced only when the effective luminosity of the disk (i.e. 
the luminosity of the disk, $L_D$ times the total line opacity 
in the optically thin case, $M_{max}$) exceeds the Eddington limit 
(e.g., Proga, Stone \& Drew 1998, PSD hereafter).
For the BH mass $M_{BH}=10^8~\MSUN$ of a typical quasar, PK found that 
for $L_D > 0.3 L_{Edd}$ a strong disk wind develops 
whereas for $L_D \simless 0.3 L_{Edd}$ there is no disk wind. 
For a less luminous disk or stronger ionizing radiation that reduces 
$M_{max}$, or both, the line force can still lift material off the disk 
but it fails to accelerate the flow to escape velocity. 
Such a failed disk wind has been
found in simulations with and without X-ray ionization (see PSD~98 for 
no X-ray cases, runs 1 and 6 there, and PSK and PK  for X-ray cases). 

For  X-ray cases, a large fraction of the failed wind is not fully ionized
and its temperature is comparable the disk effective temperature, $T_D$
(see Fig.~1). 
LD can then change the vertical structure of an accretion disk 
by dynamically increasing the disk scale height. Therefore, a failed wind 
solution can be referred to as a puffed-up disk. 

The disk wind solution is very sensitive to
$M_{BH}$: for a fixed ratio, $L_D/L_{Edd}$
(e.g., 0.5) it is easier to produce a wind for
$M_{\rm BH} \simgreat 10^7~\MSUN$ than for
$M_{\rm BH} \simless 10^7~\MSUN$ (PK). 
Thus less dramatic but very important changes in the density distribution
above the disk occur for a broad range of AGN luminosities and BH masses
(e.g., $L_D M_{max} < L_{Edd}$).
We note that LD can change the density profiles 
even inside the disk. For example, dense and massive clumps
of gas lifted by radiation can fall back on the disk (e.g., Fig. 1)
and perturb the disk structure and radiative properties.

To illustrate properties of the puffed-up disk/failed wind, we
show results from one of PK's simulations for an AGN with $M_{BH}=10^8~\MSUN$
(we refer a reader to PK for a description of the calculations).
PK present detail results from only one of the simulations
where $M_{BH}=10^8~\MSUN$ and $L_D=0.5~L_{Edd}$.
Here, we present and discuss results for the same $M_{BH}$ but
for $L_D=0.3~L_{Edd}$. For these parameters,
the disk radiation launches the flow off the disk but fails
to accelerate it to escape velocity (Fig.~1).

The top panel of Fig.~1 shows very clearly a high density
flow above the disk, i.e., the line force can maintain gas with
the density as high as $10^{-14}~{\rm g~cm^{-3}}$ as high as
$40\%$ of radius along the equator.
We note that in the radiation dominated regime,
the standard disk theory predicts
the disk scale height, $H_D=3 L/L_{Edd}~r_\ast=0.9 r_\ast$ 
(where $r_\ast\equiv 6 GM_{BH}/c^2$ is the inner disk radius and
unit of the length scale: $r'=r/r_\ast$ and $z'=z/r_\ast$).
The density distribution in the vertical
direction  is much broader compared not only to the standard disk model
but also compared to the X-ray heated disk corona. For example,
for the gas temperature $T=8\times10^8$~K, the scale height of a disk corona 
in hydrostatic equilibrium (HSE), 
$H_C/r_\ast=[k T r^3/(\mu m_p GM_{BH})]^{1/2}=0.7$ at $r'=20$.
The puffed-up disk is far from HSE and displays unsteady 
behavior: the flow is complex with a few filaments 
and various knots and clumps of gas moving both upwards and
downwards.  The direction and speed of motion at any one position is apt
to change unpredictably with time. 
There are two main reasons for this behavior: 
(i) the gravity and driving flux differently scale with $z'$ 
(e.g., PSD) and (ii) the overionization of the innermost
flow by the central engine radiation (compare the left and right panels).

PK's line-driven wind model takes into account X-ray radiation 
from a point source
located at the origin of the coordinate system. Thus it does not
account for the X-ray emission as envisioned in the disk corona scenario
described in \S2.1. However, we can estimate some of the effects of
X-rays emitted immediately above the UV disk.
For example, we can estimate the photoionization parameter, 
$\xi\equiv 4 \pi F_x/n$ inside the puffed-up disk ($F_x$ is the 
X-ray flux and $n$ is the number  density).
We assume that at a give point ($r', z'$) above the disk,
the ionizing flux  is comparable to the flux emitted by the 
disk at ($r',0$), i.e., $F_x(r',z')\approx \sigma T^4_D(r')$ (where $T_D$
is the effective disk temperature at $r'$).
For $r'=20$ this yields $T_D=20, 000$~K  and 
$F_X=2\times 10^{13}~{\rm erg~s^{-1}~cm^{-2}~s^{-1}}$.
For the density of $10^{-13}~{\rm g~cm^{-3}}$, typical for the region
above the disk at $z'\approx 4$, $\xi=5000$. 
This high $\xi$ implies that at this location, the gas would be 
fully ionized. However, for a few of reasons, 
this does not mean that the energy released above the disk
must suppress formation of the failed wind.

Firstly, the failed wind is opaque as illustrated in the middle panel 
of Fig.~1 which plots an estimate of the electron scattering optical depth
as a function of position over the unit length scale, 
$\tau\equiv \rho r_\ast \sigma_e$. 
Thus the size of a region fully ionized by 
locally dissipated energy will be small compared
to the size of the failed wind. In other words, the failed wind
can likely shield itself from locally produced X-rays as it does
from external X-rays. 
Secondly, as we eluded to above,
the presence of dense and cold gas above the disk creates unfavorable
conditions for transport and liberation of the energy 
above the disk in the first place. 
Contrary to the standard disk corona
scenario where the magnetized bubble carrying energy expands almost
freely once outside the disk, here
the bubble must rise through an extended, high density, dynamic region.
Generally, in a case of a disk with a failed disk wind,
the strength of a magnetic field needed to dominate over
the gas  and radiation pressure 
outside the disk must be orders of magnitude higher than in a case
of a bare disk.
Finally, the critical assumption about the X-ray production
is the fact that the hot gas is  tenuous
so that the main cooling mechanism is inverse Compton emission.
However, for the gas density 
in the failed wind of $>10^{-14}~{\rm g~cm^{-3}}$ 
the bremsstrahlung losses will exceed the Compton losses (e.g., eq. 1 in
Haardt \& Maraschi 1993).  Therefore,
even if the hot electrons were produced
above the disk, they will cool very efficiently by bremsstrahlung
instead of inverse Compton emission and the coronal X-ray flux
would be suppressed. 
\section{Summary and Discussion}

In the standard accretion disk theory for AGN, strong X-ray radiation
is accounted for by allowing the gravitational energy to be dissipated 
above the disk. This energy dissipation is to lead to formation of a hot disk 
corona located immediately above the disk and  inside this 
corona, UV disk photons are to be up-scattered to X-rays.  
We point out here, that for a broad range of AGN properties, radiation 
pressure due to the very same UV disk photons can drive a flow from 
the disk into the corona.

LD as a mechanism producing disk winds has been studied extensively, but LD
as a mechanism changing the density distribution above and inside the disk 
has not been given its due.
Therefore, we comment mostly on some qualitative  effects
that would be very important and relevant to AGN and other accretion disk
systems such as X-ray binaries (XBs). In particular, we use some physical 
arguments, quote and present results from simulations of line-driven disk 
winds to explore possible coupling between the X-ray and UV production 
processes.

The base for our discussion are simulations of line-driven flows from 
accretion disks studied by PK. The simulations show that 
the flows are opaque and can shield themselves not only from external X-rays 
but also from X-rays produced locally (i.e., in the region just above 
the disk where the flows are driven into). These opaque flows can lead to 
a suppression of energy dissipation above the disk. If so, the corona and 
X-rays would have to be produced interior to the UV emitting part of the disk.
This means that the UV and X-ray emitting regions do not overlap
and argues against a two-phase accretion disk model (i.e.,
against the slab/sandwich type of geometries). If this is true then
we have an inverse overionization problem: UV driven flow suppresses
X-ray production. If however, line-driven flows were suppressed 
by locally emitted X-rays, then the AGN wind would have to be launched from 
the disk exterior to the UV disk where coronal activity is negligible. 
It is possible that the intermediate situation occurs where X-rays are 
produced above the disk and the disk material is launched off the disk 
at different location/time. 
Such a double activity above the disk would lead to formation of a two-phase 
corona where high temperature plasma bubbles coexist with cooler 
and denser clumps.

The wind quenching of the hot corona raises many questions about the geometry, 
dynamics and energetics of AGN. For example, if the X-rays are quenched by 
the failed wind, what happens to the energy pumped into the gas via 
magnetic fields? Could this energy contribute to the UV continuum? Will 
the resulted UV continuum drive a flow different than that predicted by 
current models? 

To address these and other issues, one would need to perform global 
simulations of a turbulent MHD accretion disk 
with line driven flows in the radiation-dominated regime.
We anticipate that this will be possible soon
as local simulations of these type without LD
have been already preformed (e.g., Turner 2004; Hirose et al. in preparation).

There are many observational implications of the fact that the radiation
driven wind or the  failed wind can quench the X-ray production.
For example, one would expect an inverse correlation between X-ray fluxes
and winds. Perhaps, the observed anti-correlation between
the relative strength of the soft X-ray flux 
and the C~IV absorption equivalent width for QSOs 
(e.g., Brandt et al. 2000)
is due to the wind quenching rather than obscuration. The wind quenching
can also play a role in solving the overionization problem in QSOs.
The expected inverse correlation between X-ray fluxes
and winds is consistent with the fact that
BALs are not observed in X-ray sources such as Seyfert galaxies.

The wind quenching can also affect the disk reflected spectrum, 
big blue bump and iron lines.
The observed spectra are complex convolutions of the primary  
and reprocessed photons because some fraction of the X-rays are always 
intercepted by the optically thick matter and reprocessed before escaping.
Thus, to understand those spectra, it is necessary to compute very carefully 
the effects of reprocessing as many physical processes play a role (e.g., 
Ross \& Fabian 1993; {\.Z}ycki et al. 1994; Nayakshin et al. 2000). 
Future calculations of X-ray spectra should include 
the effects of line-driven flows because the flows change the key parameters
determining the spectra, i.e., the temperature and optical
depth of the scattering electrons.
These effects have at their core a coupling between X-rays 
and UV photons and therefore can help to better constrain the AGN models.

We note that the X-ray/UV coupling is unlikely to operate in the solar corona.
For this reasons among others, solar and AGN coronal activities may 
significantly differ.
However, this coupling or its variant may operate 
in XBs. For AGN and XBs, the same 
two types  of disk/corona geometries are being considered 
(e.g., Zdziarski \& Gierli{\'n}ski 2004 and Fig. 14 there).
Except for smallest radii, the bound-free and 
line opacities and radiation flux
are likely high enough for radiation pressure to 
lift dense material off the disk in some spectral/luminosity states of XBs. 
Therefore, one can expect the sphere-disk geometry in these systems because
of quenching of the disk corona.
In fact, this geometry is inferred from analysis of the photon index of 
Comptonization spectrum and frequencies of quasi periodic oscillations 
observed in some XBs
(e.g., Titarchuk \& Fiorito 2004; Titarchuk \& Shaposhnikov 2005).
It appears then that we identified a physical mechanism that 
can determine a geometry of an accretion disk and corona in 
a wide range of accreting systems.

We thank T. Kallman, G. Richards, A. R{\'o}{\.z}a{\'n}ska,
A. Siemiginowska, and J.Stone for useful discussions. 
We acknowledge support from NASA under LTSA grant NAG5-11736
and support provided by NASA through grants  HST-AR-09947.01-A
and  HST-AR-10305.05-A from the Space Telescope Science Institute, 
which is operated by the Association of Universities for Research 
in Astronomy, Inc., under NASA contract NAS5-26555.

\clearpage

\begin{figure}
\begin{picture}(180,580)
\put(180,410){\includegraphics{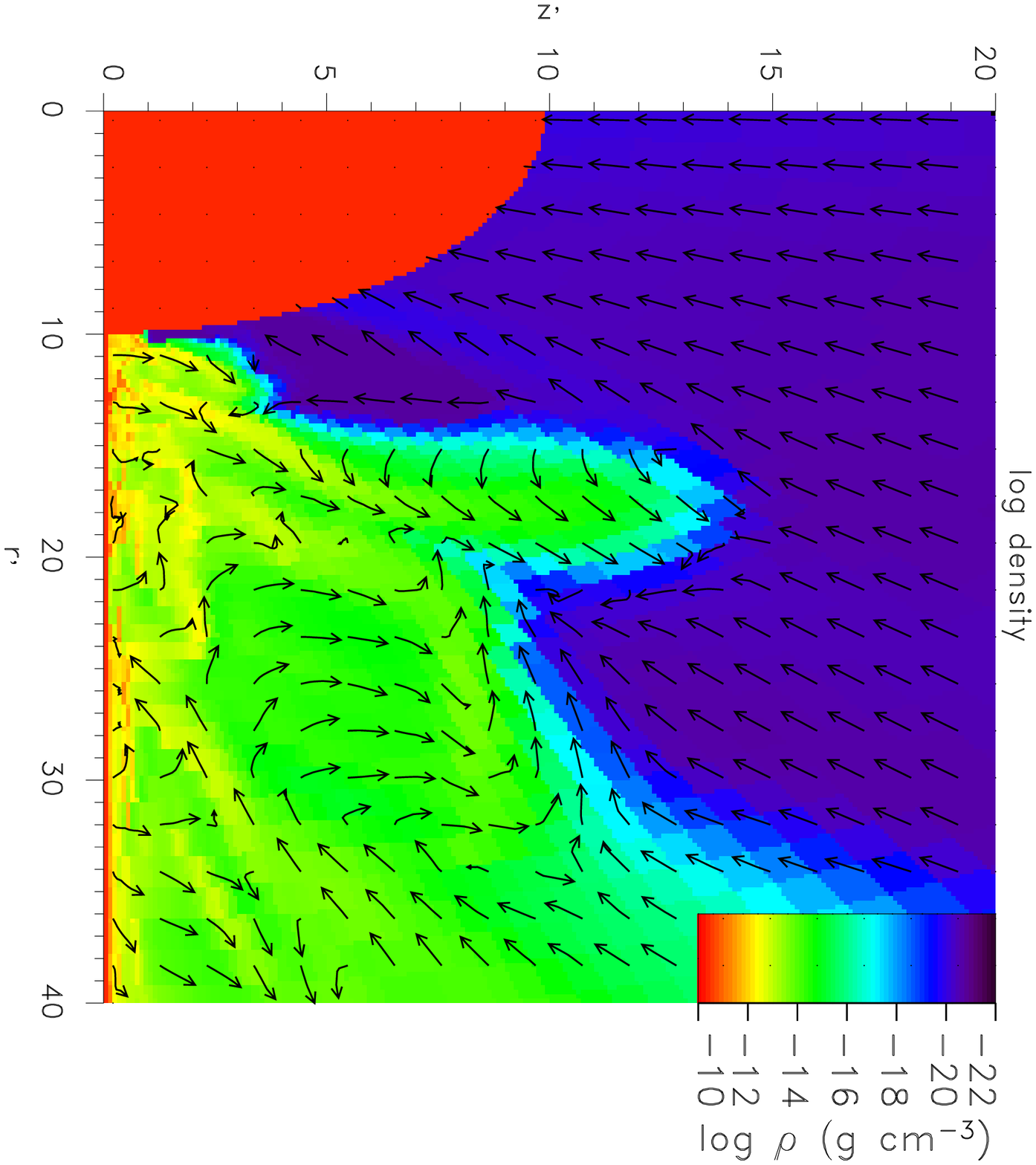}}
\put(180,210){\includegraphics{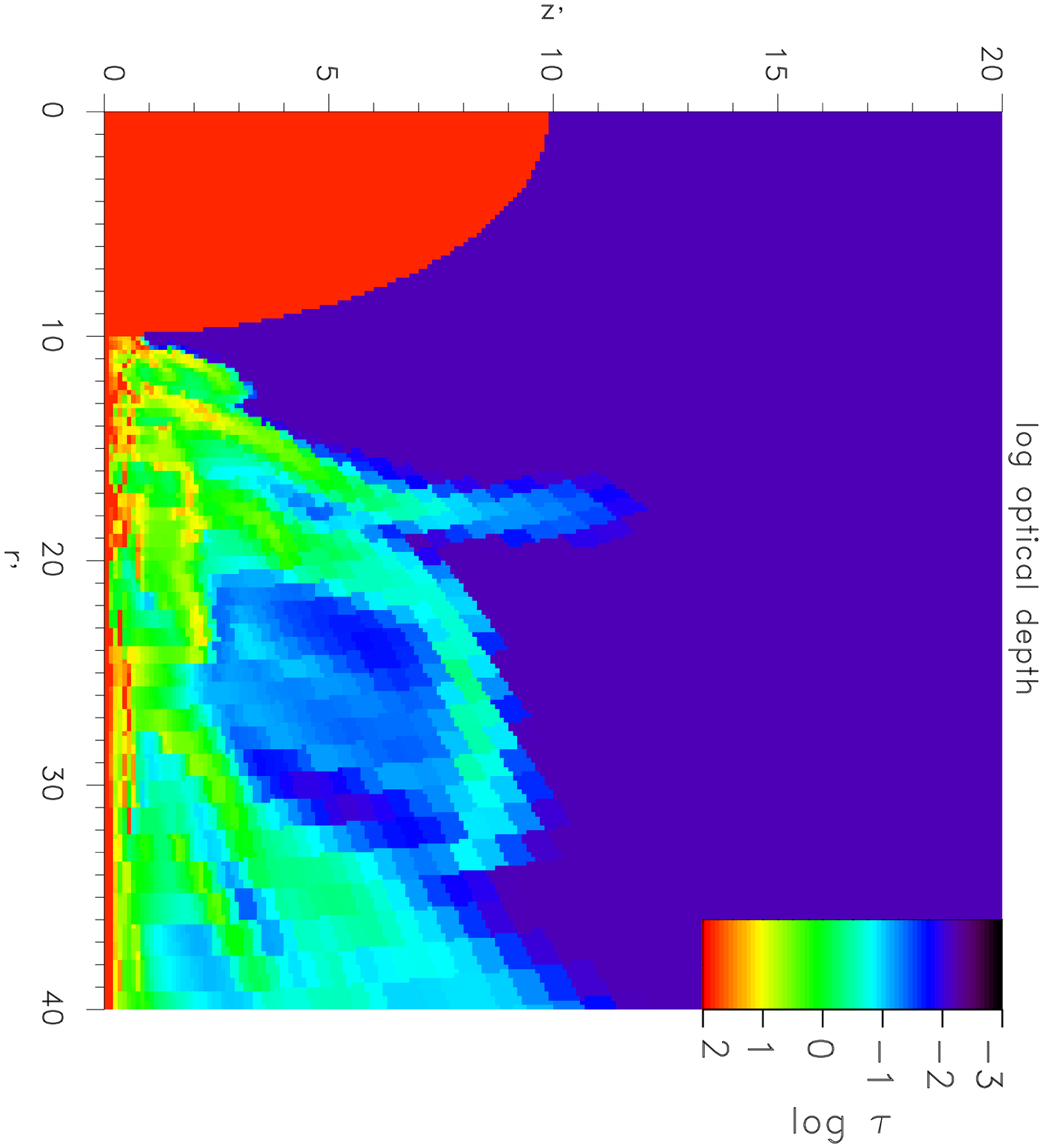}}
\put(180,10){\includegraphics{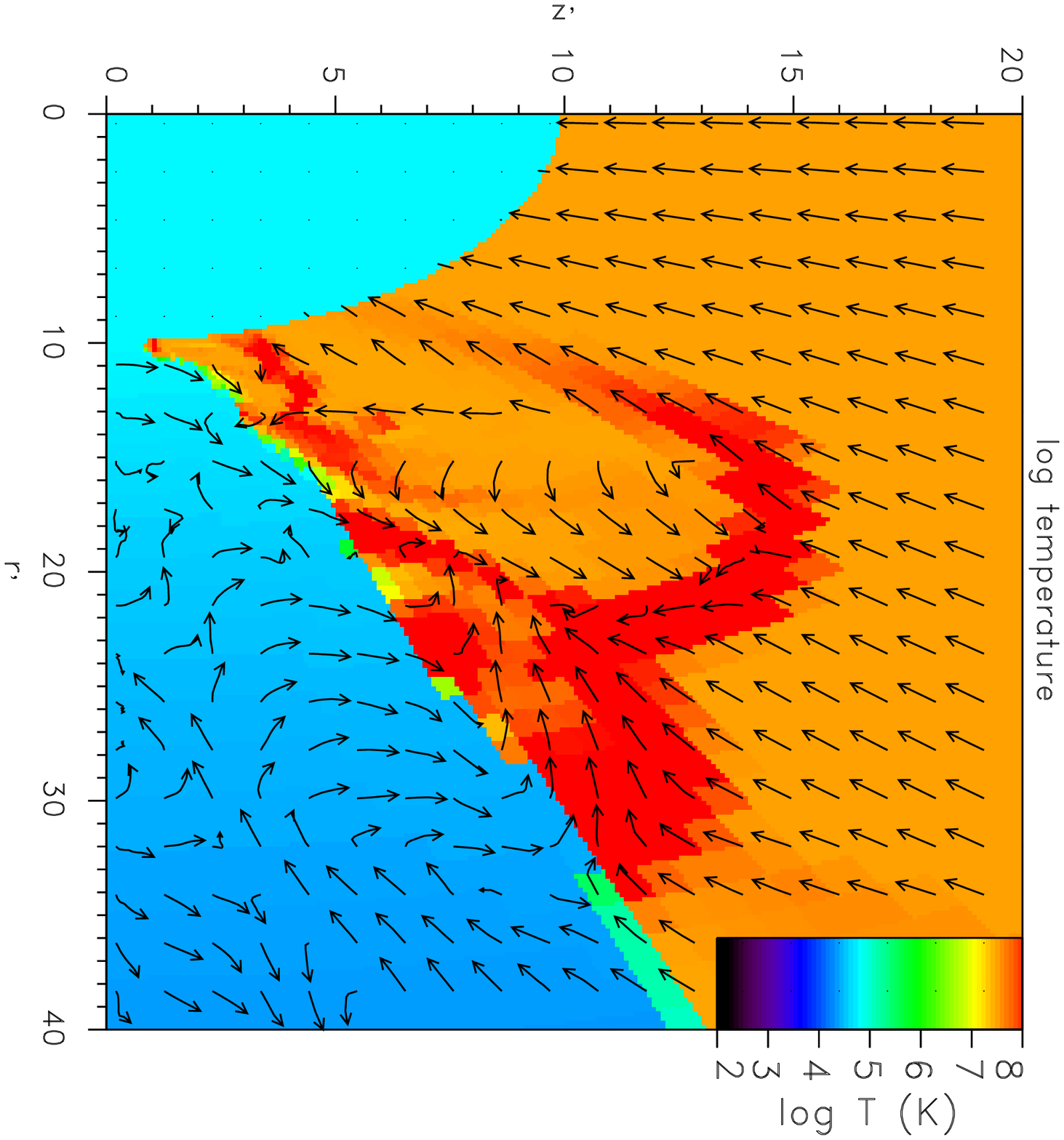}}

\end{picture}
\caption{\tiny
{\it From top to bottom:}
Maps of logarithmic density (top panel), optical depth
(middle panel), and gas temperature
(bottom panel) of the AGN failed disk wind, described in the text. 
The density and temperature maps are overplotted with
the direction of the poloidal velocity field. 
In making this figure, 
we used the density and optical depth floors of $10^{-20}~{\rm g~cm^{-3}}$
and $10^{-2}$, respectively. 
In all three panels, the disk rotational axis is along the left hand vertical
frame, while the disk photosphere is along the lower horizontal frame.
Note the difference in the range along
the $r'$ and $'z'$ axises.}
\end{figure}

\end{document}